# Real time noise and wavelength correlations in octave-spanning supercontinuum generation

T. Godin,[1,*] B. Wetzel,[1] T. Sylvestre,[1] L. Larger,[1] A. Kudlinski,[2] A. Mussot,[2] A. Ben Salem,[2] M. Zghal,[2] G. Genty,[4] F. Dias,[5] and J. M. Dudley[1]

1. Institut FEMTO-ST, UMR 6174 CNRS-Université de Franche-Comté, 25030 Besançon, France
2. PhLAM/IRCICA  CNRS-Université Lille 1, USR 3380/UMR 8523, F-59655 Villeneuve d'Ascq, France
3. University of Carthage, Engineering School of Communication of Tunis (Sup'Com), GRES'Com Laboratory, Ghazala Technopark, 2083 Ariana, Tunisia
4. Department of Physics, Tampere University of Technology, Tampere, Finland
5. School of Mathematical Sciences, University College Dublin, Belfield, Dublin 4, Ireland

* thomas.godin@femto-st.fr

**Abstract**

We use dispersive Fourier transformation to measure shot-to-shot spectral instabilities in femtosecond supercontinuum generation. We study both the onset phase of supercontinuum generation with distinct dispersive wave generation, as well as a highly-unstable supercontinuum regime spanning an octave in bandwidth.  Wavelength correlation maps allow interactions between separated spectral components to be identified, even when such interactions are not apparent in shot-to-shot or average measurements. Experimental results are interpreted using numerical simulations. Our results show the clear advantages of dispersive Fourier transformation for studying spectral noise during supercontinuum generation.

Although fiber supercontinuum (SC) generation has been the subject of extensive previous research, the SC noise properties remain a subject of much current interest [1,2]. Aside from applications such as source development, random number generation and microscopy, detailed studies of SC noise have attracted wider interdisciplinary attention because of links with extreme instabilities in other physical systems [3-6]. In this context, the development of dispersive Fourier transformation for real time spectral measurements has represented a major development, providing new insights into effects of modulation instability (MI) and rogue wave generation [7,8], as well as the statistical analysis of low power SC generation around 1.5 µm [9]. However, results to date have been limited to relatively low SC bandwidths of only around 200 nm, and the statistical data extracted from experiments has been only partially analyzed [9].

In this paper, we report two significant advances in this field of study: (i) the use of dispersive Fourier transformation (dispersive FT) to analyze shot-to-shot fluctuations of a very broadband SC spanning 550-1100 nm; and (ii) the use of a wavelength correlation map that directly shows the noise effects in particular wavelength ranges of the generated SC spectrum. The paper is organized as follows. We first describe our experimental setup and review the dispersive FT technique. We then present results from a simple numerical model to show how different types of spectral noise are manifested in the correlation maps; this will be seen to be very important to interpret the experimental results. We report experimental results in two regimes of near-infrared SC generation using a femtosecond Ti:Sapphire system: at low power where we see distinct spectral features of pump pulse nonlinear spectral broadening and dispersive wave generation; and at a higher power where we generate an octave-spanning spectrum that exhibits very significant shot-to-shot fluctuations [10]. In this latter regime, we show that the wavelength correlation map retains signatures of the underlying nonlinear spectral broadening, even though these are not apparent in the average spectrum or shot-to-shot spectral measurements. Our results clearly highlight the significant insight that can be obtained using the dispersive FT for ultrafast spectral measurements.

Fig. 1(a) shows the experimental setup. Pulses from a mode-locked Ti:Sapphire laser operating at 820 nm and with repetition rate of 80 MHz are injected via an optical isolator into a high air-fill fraction silica solid-core photonic crystal fiber (PCF). The fiber zero dispersion wavelength is ~780 nm, and nonlinearity and dispersion parameters are given below. Experiments were performed using two configurations. In the first configuration, the isolator was optimally adjusted to prevent feedback into the oscillator and near transform-limited 240fs

(FWHM) were produced, with the pulse duration calculated using autocorrelation measurements. A PCF length of 8.5 cm was used and the input power was attenuated such that the average power measured at the PCF output was 42 mW. In the second configuration, 12 cm of PCF was used and the isolator was adjusted from its optimal setting so that there was a small amount of optical feedback into the oscillator such that its operation was unstable with large intensity fluctuations and a residual continuous wave pedestal. The measured average power in this case at the PCF output was 175 mW.

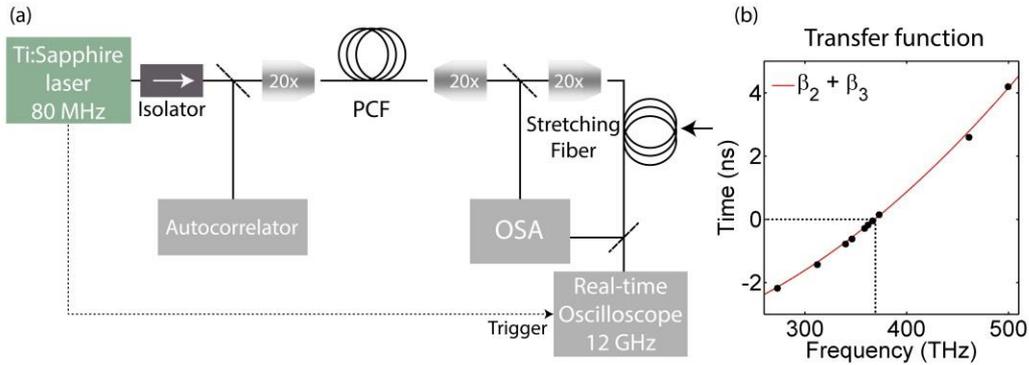

Fig. 1. (a) Experimental setup. PCF: photonic crystal fiber, OSA: Optical spectrum analyzer, (b) Frequency-to-time transfer function: Experimental measurements (circles); Fit including $\beta_2$ and $\beta_3$ (red line). Time delay is plotted relative to an origin set at the pump frequency (820 nm corresponds to 365.9 THz).

The spectrally broadened output of the PCF was characterized directly by an integrating optical spectrum analyzer (Ando AQ-6115A) as well as through dispersive Fourier transformation. The dispersive FT technique has been described in detail elsewhere [7-9] but we include a brief description here for completeness. For a temporal field $U(t)$ with Fourier transform $\tilde{U}(\omega)$, linear propagation in a length $L$ of "stretching" fiber of second–order dispersion $\beta_{2S}$ yields (for large $\beta_{2S}L$) a temporally-stretched output pulse with intensity $|U_L(t)|^2 \sim |\tilde{U}(t/\beta_{2S}L)|^2$. That is, the intensity of the output pulse yields the pulse spectrum subject to the simple mapping of the time coordinate of the stretched pulse to frequency $v$ (in Hz) where $2\pi v = t/\beta_{2S}L$. This is the principle of the dispersive Fourier transformation.

In our experiments, we performed dispersive FT spectral measurements using 100 m of custom fabricated fiber (IXFIBER IXF-SM series) specifically designed to be single-mode over a broad wavelength range in the near-infrared. The input to the stretching fiber was also attenuated to ensure linear propagation. At the pump wavelength of 820 nm, the fiber had total normal second-order dispersion $\beta_{2S}L = +4.030$ ps$^2$ and third order dispersion of $+2.344 \times 10^{-3}$ ps$^3$. Over our measurement bandwidth, the effect of third order dispersion is not negligible, but rather introduces a curvature into the time-frequency mapping. The red line in Fig. 1(b) shows the calculated nonlinear frequency-time transfer function. To confirm the calculation of this transfer function, we performed an initial series of experiments to generate SC with distinct spectral features that we could clearly identify in both the OSA spectra and the dispersive FT spectra. We were thus able to determine an experimental time delay for a range of different wavelengths. These results are shown as the circles in Fig. 1(b), confirming the dispersion parameters used in calculating the transfer function. Such a small degree of curvature in the transfer function can be readily incorporated in the time-to-frequency mapping to calibrate the frequency axis of the dispersive FT spectra [8].

We verified using the OSA to measure the SC spectra both before and after propagation in the stretching fiber that higher-order modal distortions were negligible. The typical duration of the time-stretched pulse (measured using a 25 GHz New Focus Model 1434 InGaAs photodiode) was in the range 2-6 ns at the -20 dB level in our experiments so that, with the 12.5 ns pulse period of our source laser, we could record time series of up to 32000 spectra using an extended sweep and the available oscilloscope storage capacity (Tektronix TDS6124C). Alternatively, by using a trigger at the laser repetition rate, we can display spectra repetitively at the oscilloscope refresh rate (typically 10 Hz). We checked that measurements using both approaches yielded the same SC statistical properties. The equivalent 7.5 nm spectral resolution of the dispersive FT spectral measurements was determined by the 12 GHz oscilloscope bandwidth.

Our experiments above yield a time series (ensemble) of SC spectra which can be analyzed to yield statistical moments describing the intensity fluctuations *at any given wavelength* [11]. A complementary measure of the noise properties of a SC can be obtained using the wavelength correlation map which allows correlations *between pairs of wavelengths* in the SC to be quantified [12-16], and this technique was recently applied to characterizing fluctuations over a limited SC bandwidth where the spectral broadening involved only anomalous dispersion regime dynamics [9]. Here, we consider its application in studying SC generation over broader bandwidths in both anomalous and normal dispersion regimes in the presence of dynamics involving soliton-induced dispersive wave generation.

The wavelength correlation function is defined as follows [12-15]. If $I(\lambda)$ is a time-series array of intensities at any particular wavelength $\lambda$ in the SC obtained from an ensemble of measurements, the spectral correlation between any two wavelengths $\lambda_1$ and $\lambda_2$ in the SC is given by:

$$\rho(\lambda_1, \lambda_2) = \frac{\langle I(\lambda_1)I(\lambda_2)\rangle - \langle I(\lambda_1)\rangle\langle I(\lambda_2)\rangle}{\sqrt{\left(\langle I^2(\lambda_1)\rangle - \langle I(\lambda_1)\rangle^2\right)\left(\langle I^2(\lambda_2)\rangle - \langle I(\lambda_2)\rangle^2\right)}}, \tag{1}$$

where angle brackets represent the average over the ensemble. The correlation varies over the range $-1 < \rho < 1$ with a positive correlation $\rho(\lambda_1, \lambda_2) > 0$ indicating that the intensities at the two wavelengths $\lambda_1, \lambda_2$ increase or decrease together, and a negative correlation (sometimes called anti-correlation) $\rho(\lambda_1, \lambda_2) < 0$ indicating that as the intensity at one wavelength e.g. $\lambda_1$ increases, that at $\lambda_2$ decreases and vice-versa.

Computing $\rho(\lambda_1, \lambda_2)$ for a SC ensemble yields a correlation matrix that shows the relationships between intensity variations at different wavelengths from shot-to-shot. To illustrate the utility of such spectral correlation in interpreting noisy SC generation, we first consider how several distinct types of spectral fluctuation are manifested in the correlation matrix. These results are shown in Fig. 2. As we shall see below, similar signatures will be apparent in our experimental data.

We begin by considering the spectrum $\tilde{U}(\lambda)$ of a hyperbolic secant pulse of duration 100 fs (FWHM) and bandwidth (FWHM) 7.1 nm undergoing spectral jitter around its central wavelength of the form $\tilde{U}(\lambda - \lambda_0)$. Here $\lambda_0$ is a normally-distributed random variable with mean 820 nm and standard deviation of 20 nm. Fig. 2(a) plots the correlation matrix from numerical simulations of an ensemble of 1000 spectra undergoing such wavelength jitter. The correlation plot can be understood more readily by referring its structure (the different regions of positive and negative correlation) to the spectral fluctuations shown on each axis. In these spectral plots, the grey curves show a sample of 200 realisations and the black curve is the calculated mean.

When interpreting such correlation matrix plots, we first note that we observe perfect positive correlation $\rho(\lambda_1, \lambda_2) = 1$ (yellow) across the positive diagonal when $\lambda_1 = \lambda_2$, and that the correlation is also symmetric across the positive diagonal such that $\rho(\lambda_1, \lambda_2) = \rho(\lambda_2, \lambda_1)$. In the presence of spectral jitter in Fig. 2(a), it is easy to see that different wavelengths on opposite sides of the mean will be negatively-correlated, and those on the same side of the mean will be positively correlated. This is because as jitter causes the spectrum to shift to shorter wavelengths for example, the intensity at all wavelengths shorter than the central wavelength will increase, while the intensity at wavelengths longer than the central wavelength will decrease. Thus regions $A$ and $A'$ in Fig. 2(a) which show correlations between wavelengths on the same side of the mean have $\rho(\lambda_1, \lambda_2) > 0$ whilst region B which shows correlations between opposite sides of the mean has $\rho(\lambda_1, \lambda_2) < 0$. This pattern is thus a clear signature of jitter in a pulse's central wavelength.

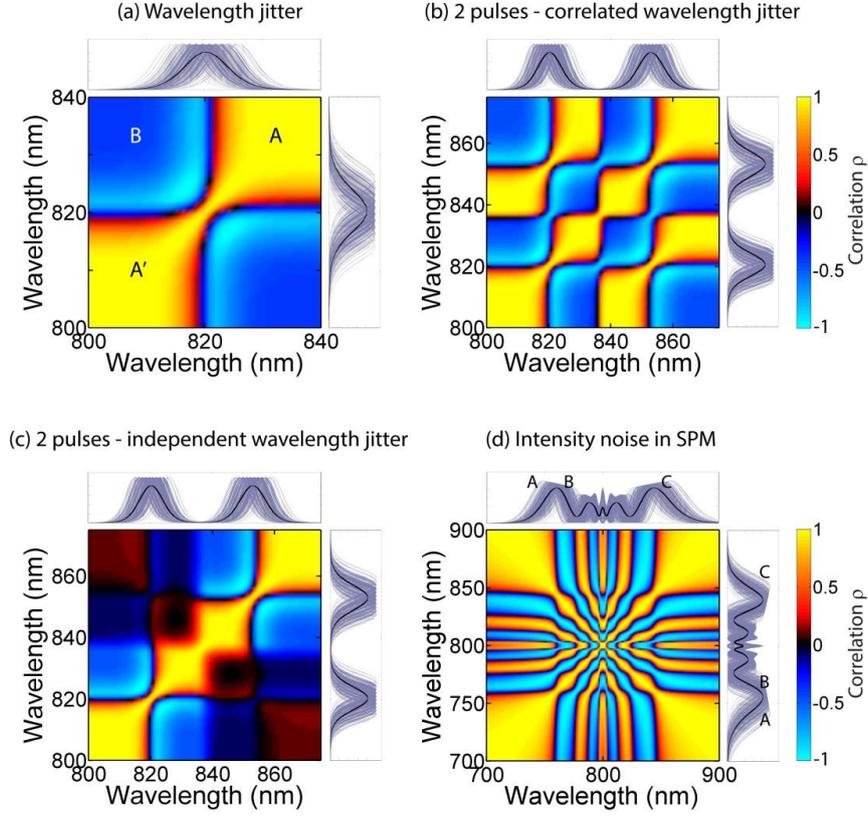

Fig. 2. Wavelength correlation maps simulated for several processes : (a) wavelength jitter; (b) wavelength jitter (two correlated pulses); (c) wavelength jitter (two independent pulses); (d) self-phase modulation (SPM) (yellow : positive correlation; blue : negative-correlation).

A similar interpretation can be made for more complex cases. Figs 2(b) and (c) show results for two pulses at different wavelengths undergoing different types of spectral jitter. Specifically, we consider a sum field $\tilde{U}(\lambda) = \tilde{U}_1(\lambda - \lambda_1) + \tilde{U}_2(\lambda - \lambda_2)$ where $\tilde{U}_1$ and $\tilde{U}_2$ are hyperbolic secant pulses as above where both central wavelengths $\lambda_1$ and $\lambda_1$ are normally-distributed random variables, with means of 820 nm and 855 nm respectively. After numerical computation of an ensemble of 1000 spectra, Fig. 2(b) shows the correlation when $\lambda_1$ and $\lambda_2$ have *identical normal distributions* such that both pulses jitter in wavelength by the same amount and in the same direction. In this case, we see positive correlation between wavelength regions on the same sides of the mean of each pulse (e.g. 800-820 nm of $\tilde{U}_1$ is correlated with 835-855 nm of $\tilde{U}_2$) and negative correlation between wavelength regions on opposite sides of the mean (e.g. 800-820 nm of $\tilde{U}_1$ is anti-correlated with 855-

875 nm of $\tilde{U}_2$). This leads to a very characteristic periodic structure of alternative regions of positive and negative correlation. In the case where $\lambda_1$ and $\lambda_2$ have *independent normal distributions* as in Fig.2(c), the periodic structure is no longer as clearly seen, and the correlation is closer to the superposition of two independent single-pulse correlation patterns as in Fig. 2(a). In general, of course, the detailed pattern observed depends on both the wavelength separation and standard deviation.

In Fig. 2(d), we consider the effect of intensity noise on nonlinear self-phase modulation (SPM). We describe SPM by $U'(t) = U(t)\exp(i\varphi_{NL}|U(t)|^2)$ where $U(t)$ is the normalized amplitude of a 100 fs hyperbolic secant pulse, and where the nonlinear phase-shift is a normally-distributed random variable of mean $\varphi_{NL} = 4\pi$ and standard deviation of $0.4\pi$. This would correspond to a pulse with large (10%) intensity noise. The effect of the noise on SPM is to cause a "breathing" of the SPM-broadened spectra as the spectral width increases or decreases for larger or smaller nonlinear phase shifts respectively. This has the effect of causing all exterior and all interior edges of the spectral peaks to be mutually correlated as their intensity will increase or decrease together, while all exterior and interior edges are mutually anti-correlated (e.g. the wavelength range in region A in the figure is correlated with region C but both A and C are anti-correlated with region B). This leads to a very characteristic periodic structure with both diagonal and horizontal and vertical zones of correlation.

We first show experimental results for the case when 240 fs pulses were used to pump an 8.5 cm length of fiber. The measured output power of 42 mW for this case allows us to estimate a peak power of 2.2 kW for the injected pump pulses, corresponding to an input soliton number of N ~ 9 and an estimated length for soliton fission of $L_{fiss}$ ~ 9 cm (using $\beta_2$ = -2.2 x $10^{-2}$ ps$^2$m$^{-1}$, $\beta_3$ = 1.2 x $10^{-4}$ ps$^3$m$^{-1}$, $\gamma$ = 50 W$^{-1}$m$^{-1}$ at 820 nm). In this regime with pumping in the anomalous dispersion regime, clear signatures of pump spectral broadening and dispersive wave generation are seen in the OSA spectrum measured at the PCF output, shown as the solid red line in Fig. 3(a). Dispersive FT measurements of the PCF output spectrum were also performed as described above, and used to record an ensemble of 1000 shot-to-shot spectra also shown in Fig. 3(a). With the high soliton number of the input pulse, the initial soliton dynamics are expected to be sensitive to input pulse noise [10], and indeed we see significant shot-to-shot fluctuations. The mean spectrum calculated from this ensemble is shown as the black line in Fig. 3(a), and there is good agreement with the OSA measurements in both the broadened pump spectral width and the spectral width and position of the dispersive wave [17,18]. Note that we do not plot data from the dispersive FT measurements over the range 660-770 nm where the signal-to-noise ratio is very low between the pump and dispersive wave and there is reduced measurement fidelity of the dispersive FT technique.

We also performed numerical simulations of spectral evolution for this regime using a stochastic generalized nonlinear Schrödinger equation model. Simulations with experimental input parameters with different initial noise were carried on with the calculated mean shown in blue in Fig 3(a) (top). There is good qualitative agreement with experiment. We also show in Fig. 3(b) the results of one individual simulation showing the pump spectral broadening and dispersive wave generation (DW) dynamics explicitly.

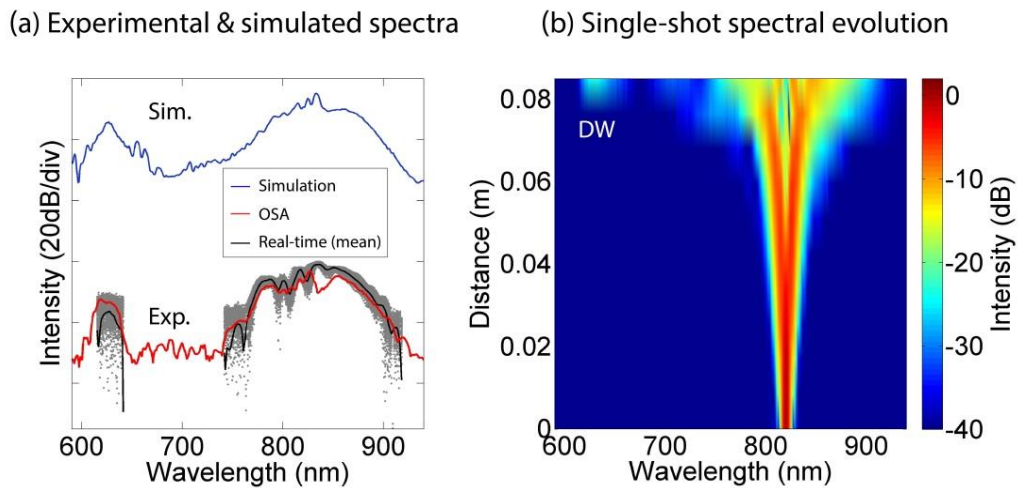

Fig.3. (a) *Top :* averaged simulated spectra (*blue line*); *Bottom* : OSA experimental spectrum *(red line);* real-time spectra (*grey dots*); averaged real-time spectrum *(black line);* (b) Spectral evolution simulated along the fiber length showing pump spectral broadening and dispersive wave (DW) generation.

From the experimental and numerical data ensemble, we can readily calculate and compare wavelength correlation maps. Figure 4 (a) plots the wavelength range of interest *around* the spectrally-broadened pump between 760-900 nm, whilst Fig. 4(b) plots the correlation *between* the broadened pump over 760-900 nm and the dispersive wave over 610-620 nm. The correlation maps show good qualitative agreement between experiment and simulation. In Fig. 4(a) we see features in the correlation maps very similar to those seen in Fig. 2(c) for specific types of spectral jitter associated with both self-phase modulation and wavelength jitter of isolated pulses. We attribute this physically to the effect of noise on the initial propagation dynamics where both spectral broadening due to self-phase modulation and the distinct pulse emergence is expected from soliton fission [16]. The correlation structure between the pump and the dispersive wave region in Fig. 4(b) is more complex but seems to indicate that spectral fringe structure around the pump is transferred to the dispersive wave region.

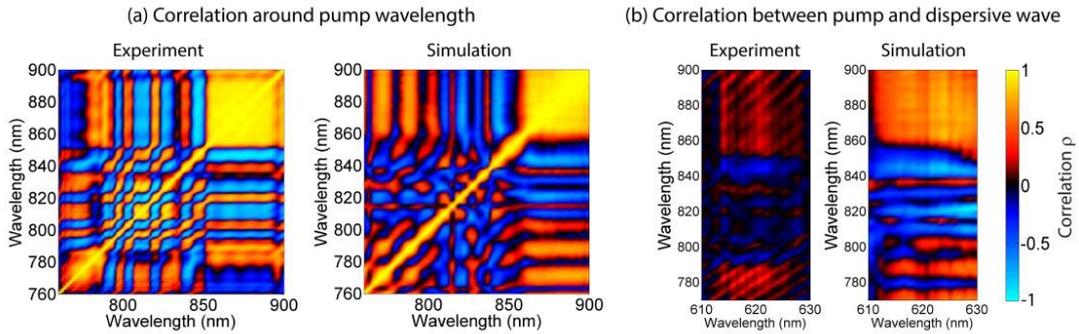

Fig. 4. Experimental and numerical wavelength correlation maps (a) Experimental correlation map around the pump wavelength (b) correlation in the pump soliton-dispersive wave region.

The second series of experimental results we show are for a highly unstable SC generated in 12 cm of PCF with feedback into the Ti:Sapphire oscillator so that we observed a strong CW peak around the pump wavelength and very large (~10%) intensity noise. The measured output average power from the PCF in this case was 175 mW but it is not possible to estimate a meaningful peak power in this case since the pulse duration could not be estimated accurately.

In these experiments, however, we readily observed SC spectra spanning from 550-1100 nm. The measured OSA spectrum is shown as the black line in Fig. 5(a). The figure also shows 1000 individual spectra measured using the dispersive FT technique (gray) as well as the calculated mean (red line); there is remarkable agreement between the OSA and dispersive FT spectra although there is reduced dynamic range in the dispersive FT results. We identify with arrows in the figure soliton peaks in the average spectrum, but we also highlight a small number of extreme red-shifted peaks (RS) that do not appear distinctly in the average spectrum but which we can identify as rogue solitons [5]. These results highlight clearly how the measurements from the dispersive FT allow us to see directly the dramatic shot-to-shot fluctuations associated with noisy SC generation; to show this more clearly, a sample of 5 individual experimental realizations is also plotted in Fig. 5(b). We note that similar plots for octave-spanning SC spectra have only previously been shown from numerical simulations and the ability to make measurements of this shot-to-shot variation is a significant experimental advance [16].

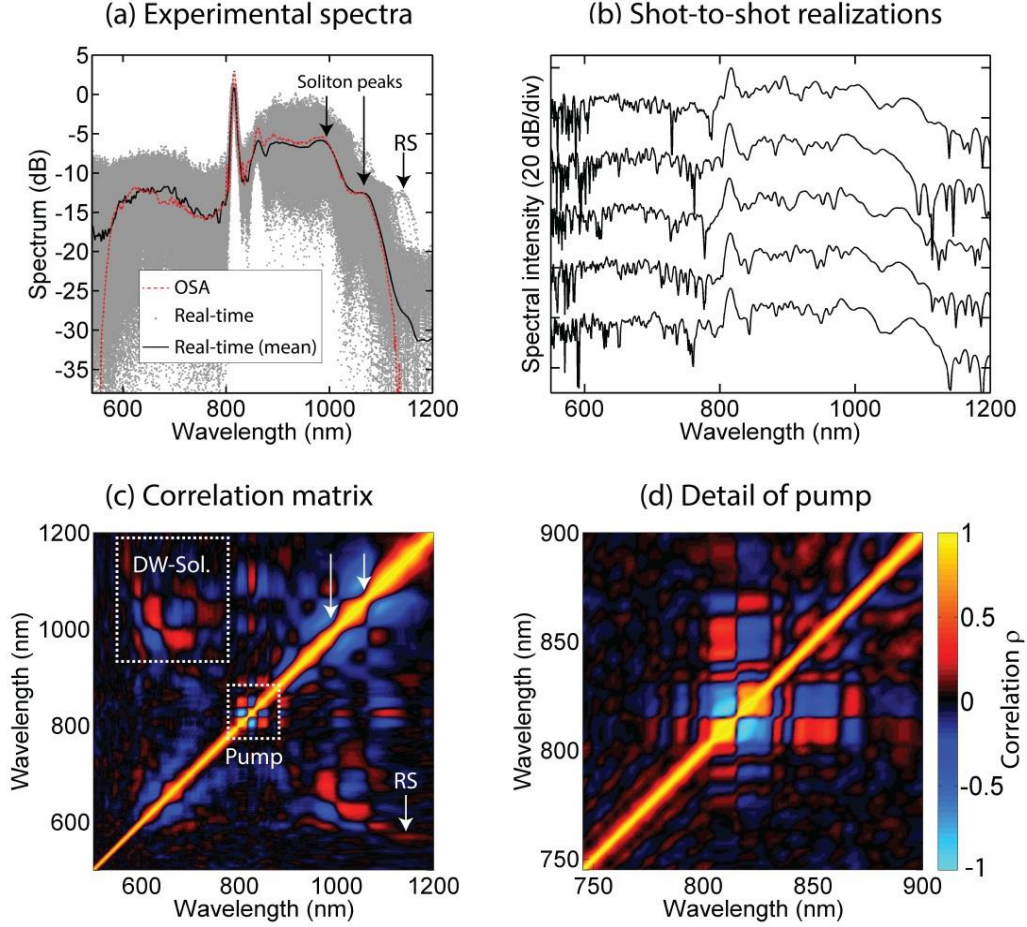

Fig. 5. Experimental results for highly unstable SC generation. (a) Spectral measurements from OSA (*red dashed line*); 1000 realisations from dispersive FT measurements showing real time fluctuations (*grey dots*); mean of these real-time spectral measurements (*black line*). The arrows show long wavelength soliton peaks and rare rogue solitons (RS). (b) A selection of 5 experimental spectral measurements using dispersive FT illustrating spectral fluctuations. (c) Wavelength correlation matrix (*DW: dispersive wave; Sol.: soliton*) (d) detail of the pump region.

Based on the ensemble measurements in Fig. 5(a), we can readily calculate the corresponding correlation matrix and Fig. 5(c) shows the calculated correlation coefficient over the full SC bandwidth. To our knowledge, the measurement bandwidth shown here is the largest for which the dispersive FT technique has been used to determine statistical data in SC generation. What is especially significant in this regime of high noise SC generation is the fact that signatures of specific nonlinear dynamics that cannot be observed from the single-shot or averaged spectra can nonetheless be seen from the correlation matrix.

In interpreting these results in Fig. 5(c), we first note that the correlation function is low almost everywhere except around the diagonal $\lambda_1 = \lambda_2$. Given the large variations in spectral structure seen in Fig 5(a) and (b) this is not surprising, but the correlation map reveals this in a particularly striking way. Nonetheless, the correlation matrix does show three particular features associated with the SC dynamics that are not apparent from the shot-to-shot spectra in Fig. 5(b). Firstly, from the averaged spectra and the physics of the SC process, we expect soliton structure on the long wavelength edge of the spectrum, with strong peaks at 990 nm and 1065 nm as indicated by arrows in Fig. 5(a). Significantly, the corresponding structure in the correlation matrix at these wavelengths (also indicated by arrows) is associated with the characteristic signature of wavelength jitter as illustrated in Fig. 2; the correlation matrix thus can show clear evidence of wavelength jitter. Secondly, although weak, we can see a complex non-zero correlation structure between the wavelength region of soliton generation (950-1050 nm) and the corresponding short-wavelength edge of the SC spectrum (650-750 nm), and we interpret this as arising from dispersive-wave soliton (DW-Sol) interactions [16]. Finally, we note characteristic features of the correlation structure around the pump wavelength region which is shown in detail in Fig. 5(d) which again shows evidence of wavelength jitter. This ability to directly identify correlated regions of the SC in the presence of noise is a major advance in our ability to study the nonlinear dynamics of SC generation and nonlinear pulse propagation.

Many fields of research can benefit from dispersive Fourier transformation in order to perform real-time measurements of ultrafast phenomena. In this paper, we have applied this technique to the study of shot-to-shot spectral instabilities in femtosecond supercontinuum generation in different regimes of spectral broadening. To our knowledge, our results represent the largest bandwidth over which dispersive FT spectral measurements have been applied to study SC noise, and our use of the wavelength correlation matrix suggests that its rich structure can be used to interpret the physics of SC interactions in a way not possible with other measurements. Although the application of dispersive FT for real time spectral characterisation does require a high speed detector and oscilloscope, these are often widely-available in laboratories focussing on experiments in ultrafast nonlinear fiber optics. We suggest that dispersive FT measurements for studies of SC noise become a routine experimental technique.

**Aknowledgements**

We acknowledge support from the European Research Council (ERC) Advanced Grant MULTIWAVE and the Agence Nationale de la Recherche (ANR OPTIROC). A. K. and A. M. acknowledge the French Ministry of Higher Education and Research, the Nord-Pas de Calais Regional Council and Fonds Européen de Développement Régional (FEDER) through the "Contrat de Projets Etat Région (CPER) 2007-2013" and the "Campus Intelligence Ambiante (CIA)".

**References**

1. J. M. Dudley and J. R. Taylor, Eds, "Supercontinuum generation in optical fibers," Cambridge University Press (2010).

2. G. P. Agrawal, "Nonlinear fiber optics," 5th edition, Academic Press, Boston (2012).

3. B. Wetzel, K. J. Blow, S. K. Turitsyn, G. Millot, L. Larger, and J. M. Dudley, "Random walks and random numbers from supercontinuum generation," Opt. Express 20, 11143-11152 (2012).

4. A. Kudlinski, B. Barviau, A. Leray, C. Spriet, L. Héliot, A. Mussot, "Control of pulse-to-pulse fluctuations in visible supercontinuum," Opt. Express 18, 27445-27454 (2010).

5. D. R. Solli, C. Ropers, P. Koonath, B. Jalali, "Optical rogue waves," Nature 450, 1054-1057 (2007).

6. B. Kibler, J. Fatome, C. Finot, G. Millot, G. Genty, B. Wetzel, N. Akhmediev, F. Dias, J. M. Dudley, "Observation of Kuznetsov-Ma soliton dynamics in optical fibre," Sci. Rep. 2, 463 (2012).

7. D. R. Solli, G. Herink, B. Jalali, C. Ropers, "Fluctuations and correlations in modulation instability," Nature Photon. 6, 463-468 (2012).

8. K. Goda and B. Jalali, "Dispersive Fourier transformation for fast continuous single-shot measurements," Nature Photon. 7, 102-112 (2013).

9. B. Wetzel, A. Stefani, L. Larger, P. A. Lacourt, J. M. Merolla, T. Sylvestre, A. Kudlinski, A. Mussot, G. Genty, F. Dias, J. M. Dudley, "Real-time full bandwidth measurement of spectral noise in supercontinuum generation," Sci. Rep. 2, 882 (2012).

10. J. M. Dudley, G. Genty, S. Coen, "Supercontinuum generation in photonic crystal fiber," Rev. Mod. Phys. 78, 1135-1184 (2006).

11. S. T. Sørensen, O. Bang, B. Wetzel, J. M. Dudley, "Describing supercontinuum noise and rogue wave statistics using higher-order moments," Opt. Commun. 285, 2451-2455 (2012).

12. E. Schmidt, L. Knöll, D. G. Welsch, M. Zielonka, F. König, A. Sizmann, "Enhanced Quantum Correlations in Bound Higher-Order Solitons," Phys. Rev. Lett. 85, 3801–3804(2000).

13. R. K. Lee, Y. C. Lai, Y. Kivshar, "Quantum correlations in soliton collisions," Phys. Rev. A 71, 035801 (2005).

14. P. Béjot, J. Kasparian, E. Salmon, R. Ackermann, N. Gisin, J. P. Wolf, "Laser noise reduction in air," Appl. Phys. Lett. 88, 251112 (2006).

15. P. Béjot, J. Kasparian, E. Salmon, R. Ackermann, J. P. Wolf, "Spectral correlation and noise reduction in laser filaments," Appl. Phys. B 87, 1–4 (2007).

16. D. Majus, A. Dubietis, "Statistical properties of ultrafast supercontinuum generated by femtosecond Gaussian and Bessel beams: a comparative study," J. Opt. Soc. Am. B 30, 994-999 (2013).

17. N. Akhmediev, M. Karlsson, "Cherenkov radiation emitted by solitons in optical fibers," Phys. Rev. A 51, 2602–2607 (1995).

18. M. Erkintalo, Y. Q. Xu, S. G. Murdoch, J. M. Dudley, and G. Genty, "Cascaded phase matching and nonlinear symmetry breaking in fiber frequency combs," Phys. Rev. Lett. 109, 223904 (2012).